\documentclass{ieeeaccess}
\usepackage{cite}
\usepackage{amsmath,amssymb,amsfonts}
\usepackage{algorithmic}
\usepackage{graphicx}
\usepackage{textcomp}
\usepackage[caption=false,font=footnotesize]{subfig} 
\usepackage{bm}
\makeatletter
\AtBeginDocument{\DeclareMathVersion{bold}
\SetSymbolFont{operators}{bold}{T1}{times}{b}{n}
\SetSymbolFont{NewLetters}{bold}{T1}{times}{b}{it}
\SetMathAlphabet{\mathrm}{bold}{T1}{times}{b}{n}
\SetMathAlphabet{\mathit}{bold}{T1}{times}{b}{it}
\SetMathAlphabet{\mathbf}{bold}{T1}{times}{b}{n}
\SetMathAlphabet{\mathtt}{bold}{OT1}{pcr}{b}{n}
\SetSymbolFont{symbols}{bold}{OMS}{cmsy}{b}{n}
\renewcommand\boldmath{\@nomath\boldmath\mathversion{bold}}}
\makeatother

\def\BibTeX{{\rm B\kern-.05em{\sc i\kern-.025em b}\kern-.08em
    T\kern-.1667em\lower.7ex\hbox{E}\kern-.125emX}}

\begin{document}
\history{Date of publication xxxx 00, 0000, date of current version xxxx 00, 0000.}
\doi{10.1109/ACCESS.2024.0429000}

\title{CSI Phase Averaging for High-Sensitivity \\ Wi-Fi Sensing in Low-Multipath Environments}
\author{\uppercase{Toshinori Suzuki}\authorrefmark{1}, \IEEEmembership{Member, IEEE},
\uppercase{Shin-ichiro OGURA}\authorrefmark{2}, 
\uppercase{Yu MORISHIMA}\authorrefmark{1},\IEEEmembership{Member, IEEE}, 
and Hiroshi MATSUURA\authorrefmark{1}
}

\address[1]{Tohoku Gakuin University, Faculty of Engineering,
3-1 Shimizukoji, Wakabayashi-ku, Sendai-shi, Miyagi 984-8588 Japan
 }
\address[2]{Tohoku University Graduate School of Agricultural Science, 
232-3 Naruko Onsen Yomogida, Osaki-shi, Miyagi, 989-6711 Japan}
\tfootnote{This work was supported by JSPS KAKENHI Grant Number JP23K03870.}

\markboth
{Suzuki \headeretal: CSI Phase Averaging for High-Sensitivity Wi-Fi Sensing}
{Suzuki \headeretal: CSI Phase Averaging for High-Sensitivity Wi-Fi Sensing}

\corresp{Corresponding author: Toshinori Suzuki (e-mail: tn.suzuki@mail.tohoku-gakuin.ac.jp).}

\begin{abstract}
This paper presents a low-complexity motion detection method for outdoor Wi-Fi sensing based on a model-driven approach. 
The method exploits the structural characteristics of the phase components in channel state information (CSI) for low-multipath propagation environments, which are generally considered disadvantageous for Wi-Fi sensing, to mitigate the phase offset errors originating from wireless devices. 
In addition, phase averaging provides a processing gain that reduces the random noise components, including quantization and thermal noise. 
The theoretical basis of the method is described and its effectiveness is experimentally evaluated using Compressed Beamforming frames obtained from commercial IEEE 802.11ac devices. 
The experiments primarily focus wild crows flying in an outdoor orchard environment. 
The experimental results demonstrate that the method can detect birds even when they fly several meters away from the direct line-of-sight path between the transmitter and receiver antennas. 
Furthermore, the results indicated that fluctuations caused by vegetation movement were negligible when the wind speed was less than 3~m/s.
The proposed approach is expected to be applicable not only to orchard monitoring but also to other outdoor Wi-Fi sensing applications in low-multipath environments. 
\end{abstract}


\titlepgskip=-21pt

\maketitle

\section{Introduction}
\label{sec:intro}
\PARstart{I}{n} recent years, crop damage caused by wild animals has become a serious problem, and effective countermeasures against bird damage in orchards and agricultural fields are  required. 
Because birds can travel over wide areas by flight, it is difficult to restrict their intrusion routes, and bird-proof nets are still widely used as a representative countermeasure. 
However, the installation and maintenance of bird-proof nets are costly, which may hinder their adoption by small-scale farms and low-profit crops \cite{A1,A3}. 
Although these studies include overseas cases and cost estimates for developing countries, the reported cost structures and economic burdens on small-scale and low income farmers have useful implications for Japanese agricultural environments, including mountainous rural regions and small family-operated farms \cite{maff2018}.

\begin{figure}[htbp]
  \centering
  \includegraphics[width=0.6\columnwidth]{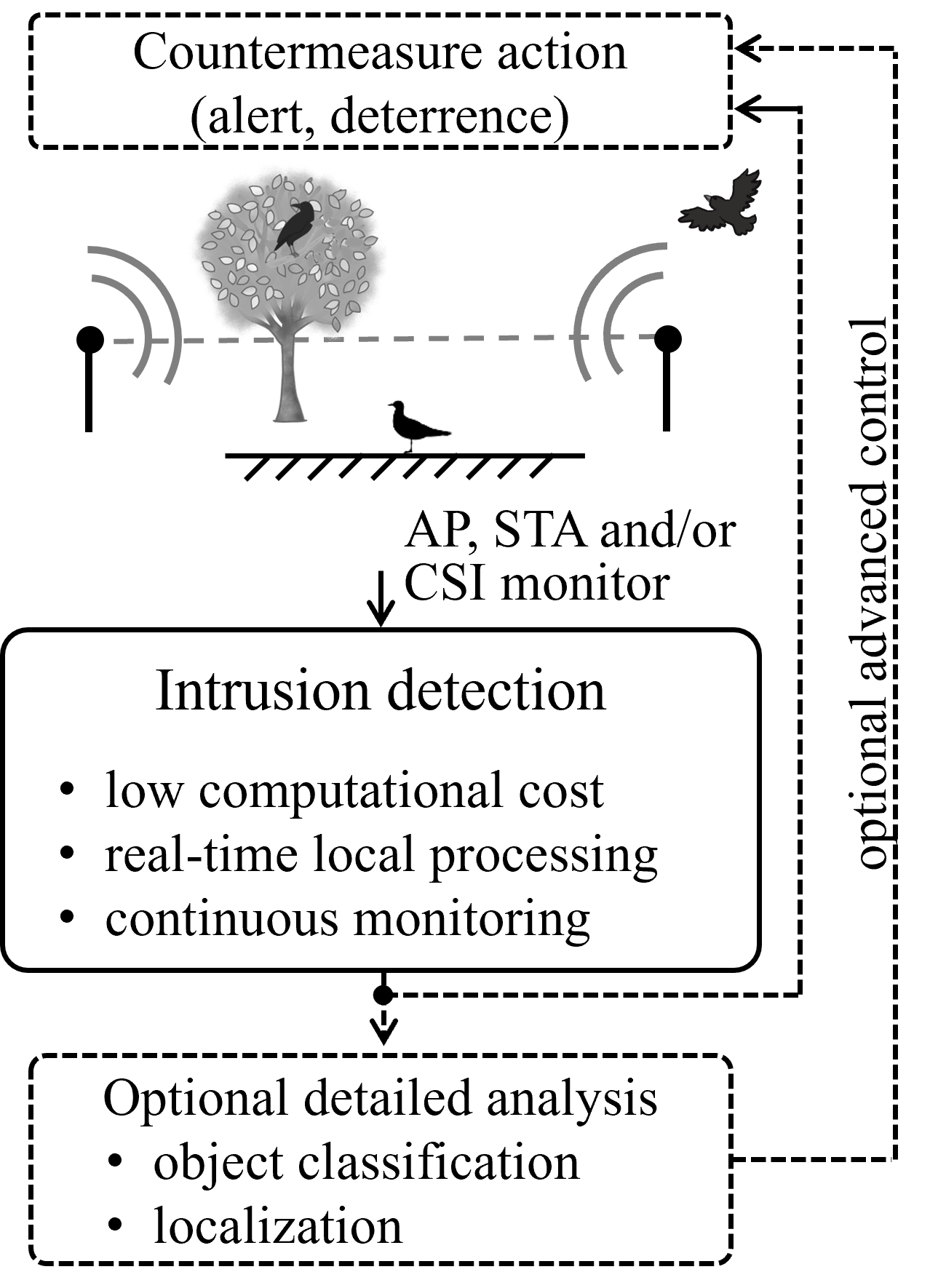}
  \caption{Outdoor Wi-Fi sensing framework for intrusion detection.}
  \label{fig:01}
\end{figure}

Therefore, the authors have been investigating bird intrusion detection in orchards using Wi-Fi sensing, which can be implemented using relatively low-cost hardware configurations. 
Early detection of approaching pest birds, followed by appropriate deterrence, is expected to contribute to damage mitigation. 
Figure~\ref{fig:01} illustrates the bird damage prevention scheme assumed in this study, and the role of intrusion detection within the overall framework. 
The objective of this study is to detect birds approaching crops as quickly as possible. 
Because such monitoring must be continuously performed, low-cost operations require local processing with reduced communication and computational overhead. 
Therefore, this study focuses on a model-based approach rather than a learning-based approach. 
In addition, the use of commercial off-the-shelf (COTS) Wi-Fi devices contributes to reducing device costs.
After detection, various countermeasures can be considered, such as deterrence using sound or lasers, as well as more advanced identification processing that enables selective deterrence based on the target type or intrusion area. 
Note that these post-detection processes are outside the scope of this study.

Conventional Wi-Fi sensing studies have primarily focused on human activity detection and behavior recognition in indoor environments \cite{guarino26,zhu24,amenta24,hernandez23,tan22,wang21,uchiyama21,he20,ma19,zhang20}. 
Research focusing on outdoor environments \cite{shimizu24,li22,miyazaki19} or non-human targets \cite{zou23,tan18} remains limited.
Reports on Wi-Fi sensing for wild bird detection remain very limited.
Compared with conventional indoor human-centered sensing scenarios, the outdoor agricultural environment considered in this study presents the following challenges.

\begin{enumerate}

\item \textit{Limited sensing range:}
In indoor environments, rich multipath reflections from walls, ceilings, furniture, and other surrounding objects enable the detection of subtle target motions even when the target is outside the line-of-sight (LoS) path, thereby extending the sensing coverage \cite{guarino26,zhou15}. 
In contrast, in low-multipath environments, target detection relies mainly on shadowing effects when the target passes near the LoS path. 
Consequently, the effective sensing range is significantly restricted \cite{koo26,li22,miyazaki19,zhou15}. 
\label{prob:air}

\item \textit{Small targets:}
Compared with humans, birds are smaller, may move more rapidly, and can intrude from above. 
Conventional Wi-Fi sensing systems are primarily designed for human-scale targets, whereas birds have a smaller radar cross section (RCS) and exhibit rapid and transient motion. 
Consequently, the channel perturbations introduced into channel state information (CSI) are weaker than those caused by humans \cite{zou23,wicpd22}. 
\label{plob:small-fast}

\item \textit{Large environmental fluctuations:}
In agricultural fields and orchard environments, the wind-induced movement of leaves and branches introduces continuous and unpredictable temporal fluctuations into both the amplitude and phase of CSI \cite{wicpd22,miyazaki19}. 
These vegetation-induced fluctuations overlap with the weak channel perturbations caused by birds, making it necessary to distinguish the actual targets from background clutter.
\label{plob:wind}

\end{enumerate}

To address Challenges \ref{prob:air}) and \ref{plob:small-fast}), it is necessary to suppress the device-induced errors contained in the measured CSI and improve the sensing sensitivity compared with conventional human-centered sensing systems. 
Furthermore, to address the rapid motion of birds associated with Challenge \ref{plob:small-fast}) and the environmental fluctuations described in Challenge \ref{plob:wind}), it is desirable to focus on high-frequency CSI variations rather than low frequency components.

A model-based Wi-Fi sensing approach generally consists of three elements: CSI collection, CSI-preprocessing, and feature extraction based on the model \cite{wang21}. 
Various physical models and signal processing models have been employed as the basis of conventional Wi-Fi sensing systems, including the angle of arrival (AoA) model, Fresnel-zone model, CSI-speed model, and Doppler model \cite{wang21,li24}. 
Moving variance segmentation (MVS), which evaluates CSI fluctuations within a specified time window, is one of the most representative approaches for motion detection \cite{wang22,wang21}. 
The squared amplitude of the CSI corresponds to the received signal power and therefore has a clear physical interpretation.

\begin{figure}[h]
 \centering
 \subfloat[conventional model base]{
 \includegraphics[height=5.5cm]{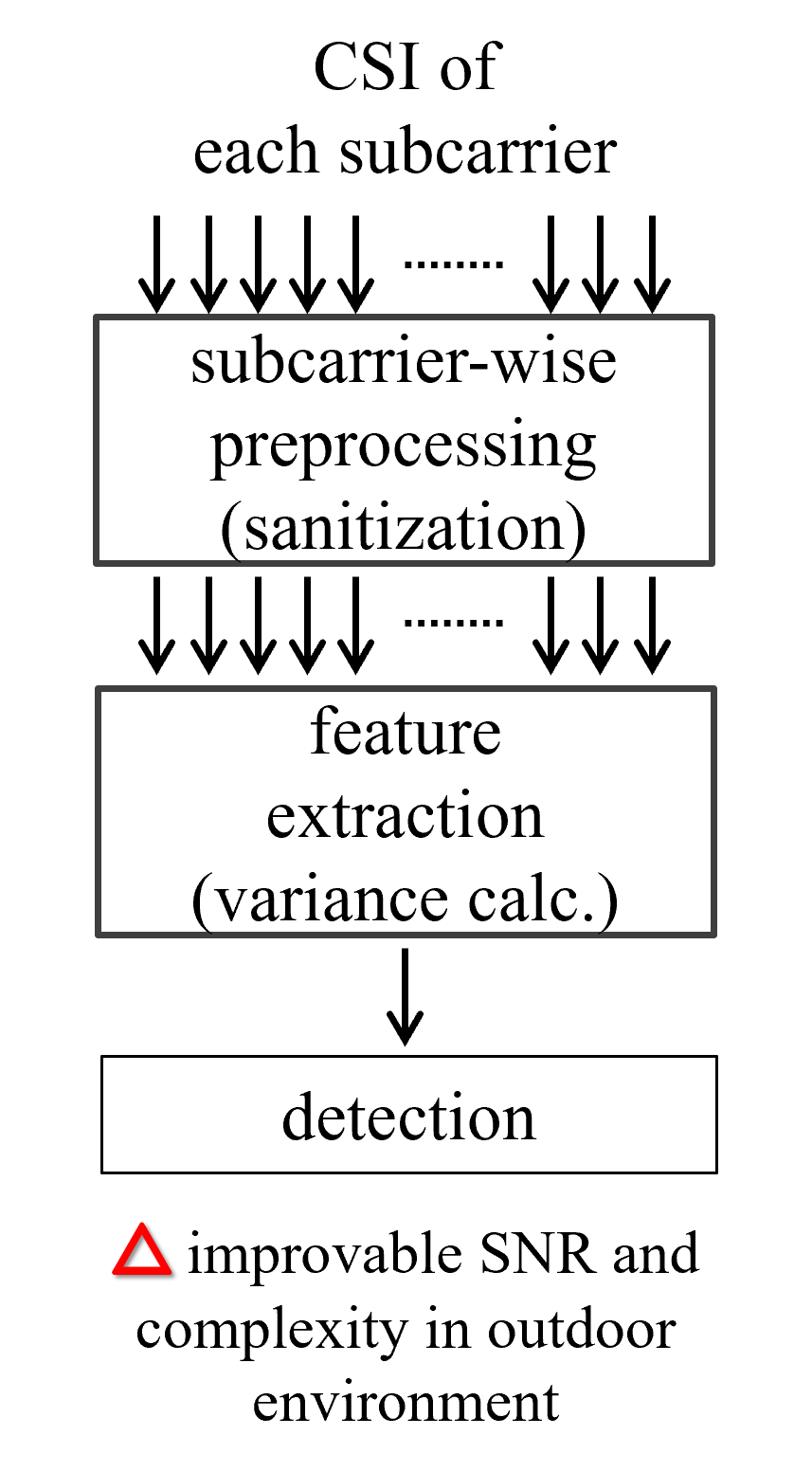}
 \label{fig:02a}
 }
 \hspace{4mm}
 \subfloat[low-multipath model base]{
 \includegraphics[height=5.5cm]{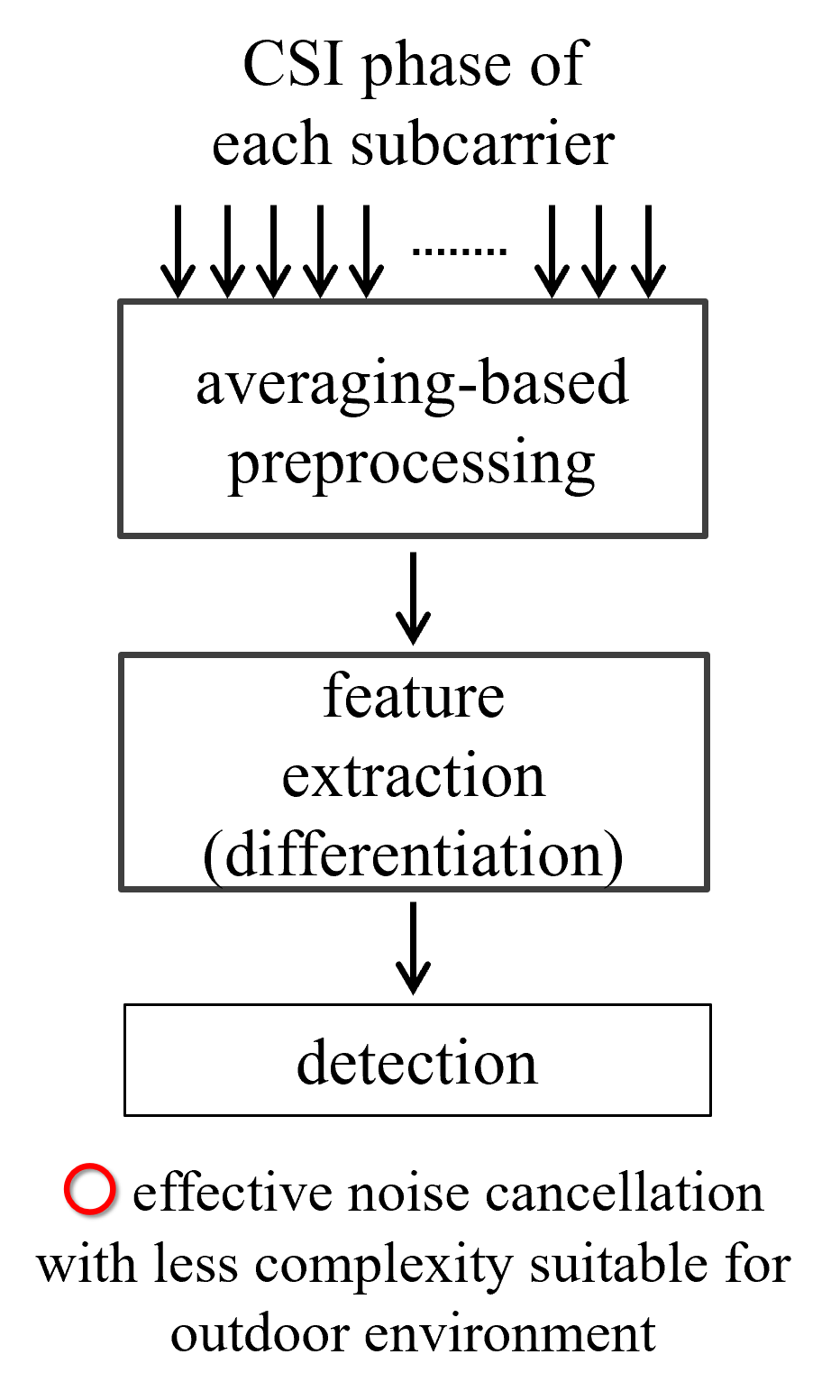}
 \label{fig:02b}
 }
 \caption{Comparison of CSI processing methods for motion detection.}
 \label{fig:02}
\end{figure}

A typical processing flow based on these conventional approaches is illustrated in Fig.~\ref{fig:02}(a). 
The device-induced errors in the measured CSI are first reduced individually for each subcarrier.
Motion-related features, such as the MVS, are then extracted from the CSI variations and integrated to determine the presence of moving objects.

In contrast, the proposed method exploits the fact that in low-multipath environments traditionally regarded as unfavorable for Wi-Fi sensing, the structure of the true CSI phase across subcarriers becomes similar to that of device-induced phase offsets.
This property enables phase averaging to suppress both device-induced errors and random noise without regression-based sanitization.
Based on this characteristic, the proposed method simultaneously achieves a low computational complexity and high sensing sensitivity. 
Specifically, as shown in Fig.~\ref{fig:02}(b), the measured CSI phase is averaged across the subcarriers during preprocessing and condensed into a single representative value. 
The motion is then detected based on its temporal variation. 

Although the resulting structure is extremely simple, the proposed method not only suppresses phase offsets caused by device-induced errors, but also provides a processing gain that reduces random noise components, including quantization and thermal noise, which are not explicitly addressed by conventional models. 
This paper presents the theoretical basis of the proposed method and demonstrates its effectiveness by using CSI data measured in an outdoor environment.
Experimental results further demonstrate that the proposed method can detect bird motion even outside the direct LoS path.

Although the orchard monitoring considered in this study represents an example of outdoor Wi-Fi sensing, the proposed low-complexity motion detection method is also expected to be applicable to other outdoor Wi-Fi sensing applications such as outdoor monitoring and intrusion detection in low-multipath environments. 
Furthermore, similar low-multipath conditions may arise in communication environments employing directional antennas and in future Integrated Sensing and Communication (ISAC) systems. 
Therefore, the findings of this study are expected to provide useful insights into CSI-based sensing in low-multipath environments.

The remainder of this paper is organized as follows.
Section~\ref{sec:me} describes the outdoor measurement environment and discusses environmental fluctuations caused by wind-induced vegetation movement. 
Section~\ref{sec:proposed} presents the theoretical background of the proposed method. 
Section~\ref{sec:eval} evaluates the effectiveness of the proposed method using the CSI data measured in the environment described in Section~\ref{sec:me}. 
Finally, Section~\ref{sec:conclude} concludes this paper.

\section{Measurement Environment}
\label{sec:me}

\subsection{Measurement Setup}
\label{ssec:ms}
One access point (AP) and two stations (STAs) were installed in waterproof acrylonitrile butadiene styrene (ABS) resin enclosures in a blueberry orchard located at the Field Science Center, Graduate School of Agricultural Science, Tohoku University \cite{kawatabiFC}, as shown in Fig.~3. 
\begin{figure}[htbp]
  \centering
  \includegraphics[width=\columnwidth]{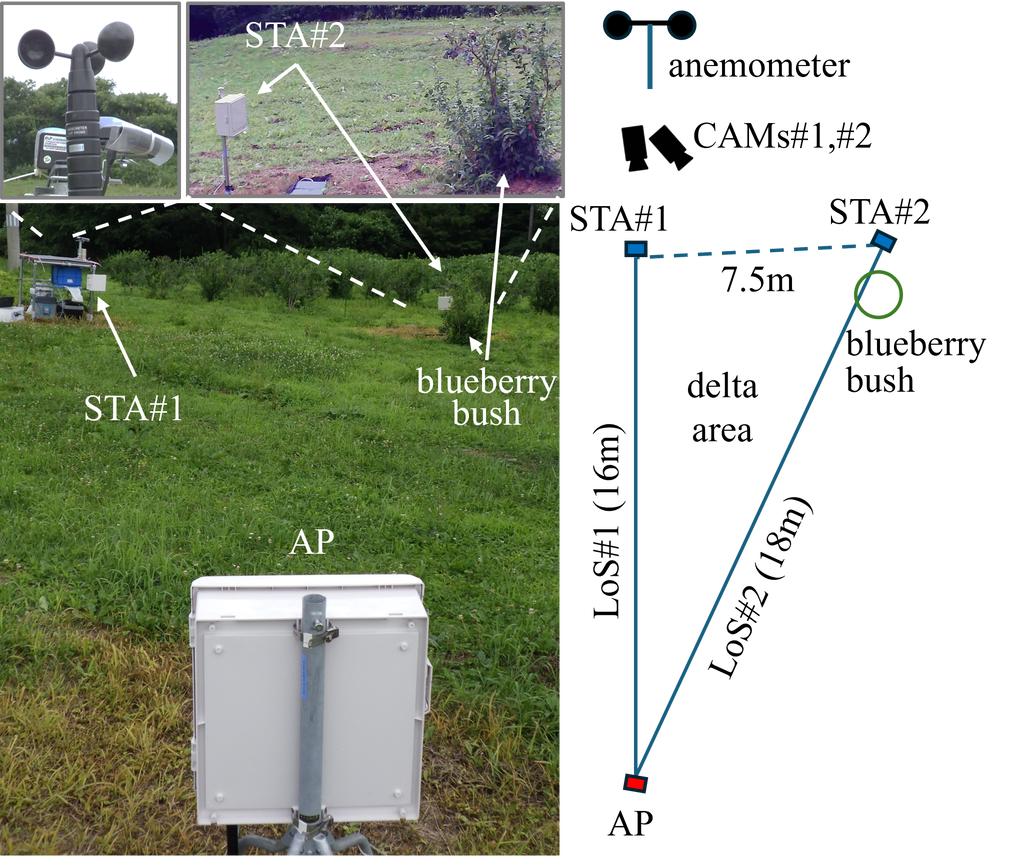}
  \caption{Measurement Environment.}
  \label{fig:03}
\end{figure}
Data collection system including a Wi-Fi frame receiver, two cameras, an anemometer, and a measurement PC were installed near STA\#1. 
As commercial power was unavailable in the orchard, the system was operated using solar power generation. 
STA\#1 was positioned to overlook the area where crows were expected to fly, whereas STA\#2 was installed near the blueberry bush to monitor bird intrusion into the orchard. 
The system specifications are listed in Table~\ref{tab:spec}.

\begin{table}[htbp]
\caption{System Specifications and Measurement Conditions}
\label{tab:spec}
\centering
\begin{tabular}{|l|l|}
\hline
\textbf{Item} & \textbf{Specification / Condition} \\ \hline
AP & Buffalo WXR-1900DHP, height: 70 cm \\ \hline
STA & Raspberry Pi CM4, height: 80 cm \\ \hline
Specification & IEEE 802.11ac \\ \hline
MIMO & $3 \times 1$ (AP $\times$ STA) \\ \hline
Channel & 5.6 GHz band (bandwidth: 20 MHz) \\ \hline
\end{tabular}
\end{table}

UDP traffic of several Mbps was generated between the AP and STAs using \textit{iperf}, and VHT Compressed Beamforming frames transmitted from the STAs were collected by the measurement PC. 
The measured data and recorded videos were timestamped based on the system clock of the PC. 
In this study, data measured between 08:00 and 13:00 JST on July 16, 2025, during which crop damage caused by crows was confirmed twice, are analyzed.

The Kawatabi weather observation station, which is part of the nationwide automated weather observation network operated by the Japan Meteorological Agency (JMA) \cite{jma},is located approximately 200~m from the measured blueberry orchard and provides publicly available weather data \cite{amedas} such as precipitation.
The corresponding weather conditions are listed in Table~\ref{tab:weather}. 
Light rainfall was observed both before and during the measurements (0.5~mm between 09:50 and 10:00 according to the 10-minute observation records).
\begin{table}[htbp]
\caption{Weather Conditions at Kawatabi very one hour on July 16, 2025.}
\label{tab:weather}
\centering
\begin{tabular}{|c|c|c|}
\hline
\textbf{Time Range} &
\textbf{Precipitation} &
\textbf{Sunshine} \\
hh:mm&
\textbf{(mm)} &
\textbf{Duration (h)} \\
\hline

-- 01:00 & 2.0 & Before sunrise \\ \hline
-- 02:00 & 1.5 & Before sunrise \\ \hline
-- 03:00 & 0   & Before sunrise  \\ \hline
-- 04:00 & 0   & Before sunrise  \\ \hline
-- 05:00 & 0   & 0  \\ \hline
-- 06:00 & 1.0 & 0  \\ \hline
-- 07:00 & 0   & 0  \\ \hline
-- 08:00 & 0   & 0  \\ \hline
-- 09:00 & 0   & 0.2  \\ \hline
-- 10:00 & 0.5 & 0  \\ \hline
-- 11:00 & 0   & 0.5  \\ \hline
-- 12:00 & 0   & 0.8  \\ \hline
-- 13:00 & 0   & 0.6  \\ \hline

\end{tabular}
\end{table}

\subsection{Compressed CSI in the Measurement System} 
\label{ssec:ccsi}

In the employed 3 $\times$ 1 MIMO configuration, three CSI components are available for each subcarrier.
Let these complex CSI values be denoted by $h_1$, $h_2$, and $h_3$. 
At the STA, the received CSI is first normalized such that $|h_1|^2 + |h_2|^2 + |h_3|^2 = 1$, and is then converted (compressed) into four angular parameters, $\varphi_{11}$, $\varphi_{21}$, $\psi_{21}$, and $\psi_{31}$, which are reported to the AP \cite{11ac}. 
The parameters $\varphi$s are quantized using six bits over the range from $0$ to $2\pi$, whereas the parameters $\psi$s are quantized using four bits over the range from $0$ to $\pi/2$. 
For the channel configuration listed in Table~\ref{tab:spec}, four angular parameters (20 bits) are reported for each of the 52 subcarriers, resulting in 1040 bits of CSI information per VHT Compressed Beamforming frame \cite{11ac}. 
The relationship between the three complex variables and four angular parameters is given by (1).

\begin{equation}
\begin{aligned}
|h_1| &= \cos \psi_{21} \cos \psi_{31}, \\
|h_2| &= \sin \psi_{21} \cos \psi_{31}, \\
|h_3| &= \sin \psi_{31}, \\
\varphi_{11} &= \angle h_3 - \angle h_1, \\
\varphi_{21} &= \angle h_3 - \angle h_2.
\end{aligned}
\label{eq:comp_csi}
\end{equation}

The original complex CSI values $h_1$, $h_2$, and $h_3$ could not be uniquely reconstructed from the four angular parameters. 
For notational convenience, the variables $\varphi_{11}$ and $\varphi_{21}$ appearing in (1) are not reused in the following discussion, unless otherwise stated, and the symbol $\varphi$ is subsequently used with a different definition.

\subsection{Can Bird Intrusion Be Distinguished From Vegetation Movement?}
\label{ssec:bird-speed}
\begin{figure}[htbp]
  \centering
  \includegraphics[width=0.7\columnwidth]{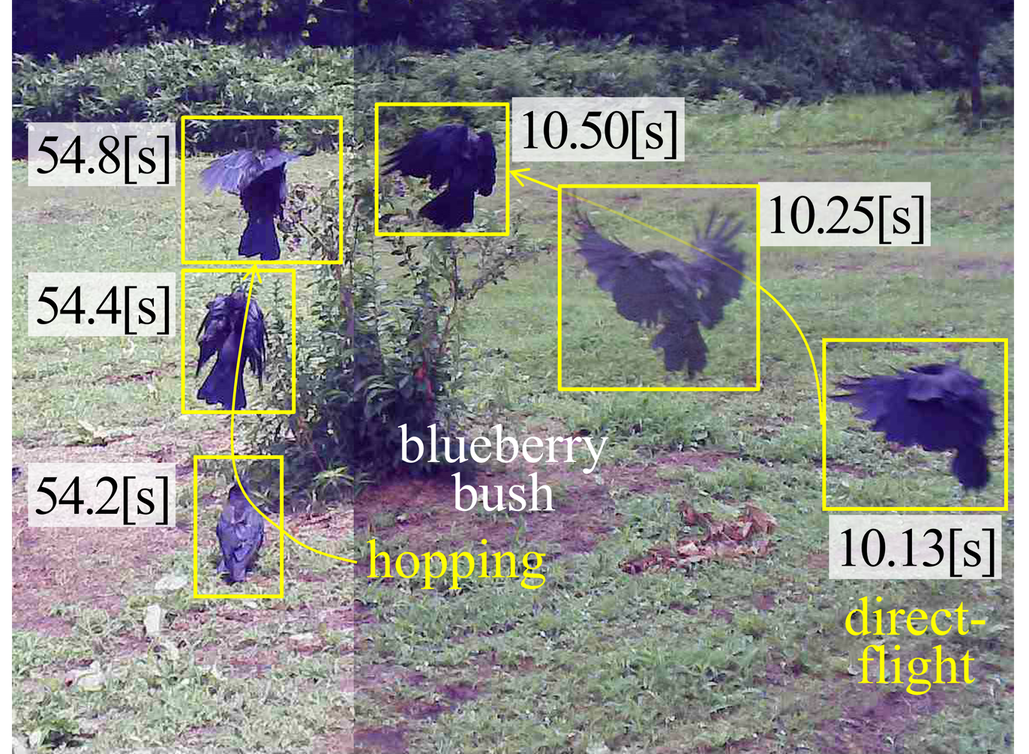}
  \caption{Examples of crow hopping and direct flight intrusion.}
  \label{fig:04}
\end{figure}

Using the recorded video data, the flight velocity of the crows immediately before landing on the blueberry bush was estimated. 
Figure~\ref{fig:04} shows the overlaid images of crows just before landing, together with the timestamps of the corresponding video frames. 
Two typical intrusion patterns were observed: hopping from the ground onto the bush (left side of Fig.~\ref{fig:04}, from 54.2(s) to 54.8(s)) and direct flight intrusion (right side of Fig.~\ref{fig:04}, from 10.13(s) to 10.50(s)). 
By estimating the movement distance using the approximate height of the blueberry bush (approximately 1.3~m), the bird velocities were estimated to be approximately 3.5~m/s and 4.8~m/s, respectively.
In contrast, the measured wind speed from 08:00 to 13:00 on July 16 ranged from 0 to 4.3~m/s, with the 90th-percentile and 99.5th-percentile values being 2.3~m/s and 3.4~m/s, respectively. 
Because the movement speed of vegetation caused by wind cannot exceed the wind speed itself, the false positive rate becomes less than 0.5\% if bird intrusion and vegetation movement can be discriminated based solely on a motion-speed threshold of 3.5~m/s.

\section{Proposed Motion Detection Method}
\label{sec:proposed}
\subsection{Basic Principle of Motion Detection}
Model-based motion detection is performed by evaluating temporal variations in the measured CSI and determining the presence of moving objects from the magnitude of the variations.
In conventional approaches, CSI variations are calculated independently for each subcarrier, and feature values such as variance or squared temporal difference are integrated across subcarriers for motion detection. 
The errors contained in CSI can be reduced using the methods described below.

\subsection{Conventional Device Error Suppression Methods}

Amplitude errors primarily originate from RF circuit imperfections, including the frequency responses of antennas and filters, temperature-dependent variations, and Automatic Gain Control (AGC) mechanisms \cite{zhang22}. 
In contrast, dominant phase errors are mainly caused by local oscillator (LO) instability and internal clock offsets \cite{zhang22,uchiyama21}.
These phase errors can be mathematically modeled as a combination of a subcarrier-independent bias component and subcarrier dependent linear component proportional to the subcarrier index.
The former is largely attributed to the Carrier Frequency Offset (CFO), while the latter is associated with the Sampling Frequency Offset (SFO) and Packet Detection Delay (PDD).
Owing to the hardware architecture of commercial Wi-Fi NICs, these clock-related phase errors are typically shared across all antennas on the same card, making them antenna-common quantities \cite{wang22,he20,wu20,zeng19,jiang22}.
Conversely, independent phase errors may still arise in each RF chain owing to variations in RF cable lengths and specific filter phase characteristics.

Noise components affecting both the amplitude and phase include thermal noise from analog circuits and quantization noise introduced by analog-to-digital converters (ADCs).
Both types of noise degrade the sensing performance, particularly as the signal strength decreases \cite{koo26,zhang22}.

To mitigate phase errors, the conventional approach utilizes linear regression (or linear fitting) to estimate the phase component proportional to the subcarrier index and subtracts it from the measured CSI phase.
Although this method can remove error components associated with timing offsets, its accuracy often degrades owing to nonlinear channel phase responses and failures in phase unwrapping, making stable error suppression challenging \cite{wang19}.

Another widely adopted phase-correction method involves calculating the phase difference relative to a reference antenna \cite{uchiyama21}.
In this spatial differencing approach, phase offsets shared across antennas (such as the CFO) are canceled out. 
However, the RF-chain-dependent phase errors specific to each antenna path remain uncompensated \cite{he20}.

This concept is further extended to the complex CSI ratio (or CSI quotient), calculated by dividing the measured complex CSI of one antenna by that of the reference antenna \cite{zeng19}.
This method compensates for antenna-common amplitude scaling (e.g., from AGC) while simultaneously suppressing shared clock-related phase offsets \cite{wang22,wu20}.
Because the estimation accuracy deteriorates when low-SNR subcarriers are used as the reference (denominator), such subcarriers are typically excluded from processing \cite{wang22,ma19}.

To mitigate residual noise components such as thermal noise that cannot be directly compensated, techniques such as SNR-based CSI weighting, principal component analysis (PCA)-based CSI denoising scheme and subcarrier selection have been investigated  \cite{hernandez23, wang15}.
Averaging highly correlated subcarriers can improve the effective SNR. 
Although such averaging has often been introduced to reduce the computational complexity during feature extraction \cite{dong18}, it also serves as an effective noise reduction strategy.

In addition to the methods described above, outlier rejection (e.g., Hampel filtering) and various other filtering techniques (e.g., low-pass or Savitzky-Golay filters) are frequently employed to eliminate high-frequency noise components \cite{ma19}.

\subsection{Exclusion of CSI Amplitude Components}

Because RF characteristics exhibit device-specific similarities across antennas, amplitude variations caused by common-mode components can be partially suppressed by using amplitude ratios. 
However, some error components contained in the CSI amplitude are independently generated in each RF chain, making complete error removal difficult when only the amplitude ratios are used. 
In particular, the accurate compensation of hardware-induced errors, such as RF gain mismatch and IQ imbalance, often requires prior calibration or additional measurements.

Furthermore, in the compressed CSI used in this study, the amplitude is normalized for each subcarrier, making straightforward SNR-based subcarrier selection difficult. 
In addition, as described in Section~\ref{ssec:ccsi}, the amplitude components are represented using a relatively small number of quantization bits. 
Because highly accurate suppression of device-induced noise is required for detecting fast-moving targets, this study focuses exclusively on CSI phase components and excludes CSI amplitude information from the proposed motion-detection framework. 

Nevertheless, the potential utilization of CSI amplitude information in low-multipath environments remains an interesting topic for future investigation.

\subsection{Structural Similarity Between True CSI Phase and Device-Induced Phase Offsets}
To suppress hardware-induced CSI phase errors, including antenna-dependent components, antenna phase differencing alone is insufficient, and the removal of phase components proportional to the subcarrier index is effective. 
However, as previously described, this approach suffers from challenges in terms of estimation stability and tracking performance.

To address this problem, low-multipath environments, which are generally regarded as disadvantageous for Wi-Fi sensing, can be beneficial. 
This is because the true CSI phase itself in such environments exhibits a frequency-dependent structure similar to that of hardware-induced phase errors.
By fully exploiting this characteristic, an efficient and accurate phase error suppression method can be derived under the assumption of low-multipath propagation environments.

Specifically, the proposed approach eliminates the need for regression analysis, thereby simplifying the processing procedure while simultaneously suppressing phase errors caused by quantization noise and thermal noise. 
Details of the proposed method are described in the following subsections.

\subsection{Averaging of Symmetric Subcarrier Phases}

Let the frequency-dependent component of the device-induced phase error for antenna \#$i$ and subcarrier \#$k$ be denoted by \[ \Phi_{i,k} = \Phi_i + k\Delta\Phi_i . \]
Then, the measured CSI phase is expressed as
\begin{equation}
\begin{aligned}
\varphi_{i,k} &= \left(\Phi_{i,k} + \varphi^{\mathrm{Air}}_{i,k} \right) \bmod 2\pi \\
&= \left(\Phi_i + k \Delta \Phi_i + \varphi^{\mathrm{Air}}_{i,k} \right) \bmod 2\pi ,
\end{aligned}
\label{eq:phi_ik}
\end{equation}
where $\varphi^{\mathrm{Air}}_{i,k}$ is the true CSI phase. 
The measured phase $\varphi_{i,k}$ takes values in the range of $-\pi$ to $\pi$ owing to the modulo operation. 
Subcarrier index $k$ is assumed to be symmetrically assigned around zero. 
Because subcarrier \#0 is null, the center phase $\varphi_{i,0}$ cannot be measured directly.

When only the direct wave is present and its path length is denoted by $l_i$, the true CSI phase can be written as
\begin{equation}
\begin{aligned}
\varphi^{\mathrm{Air}}_{i,k} = -2\pi \frac{l_i}{\lambda_k} = -2\pi (f_0 + k\Delta f)\tau_i ,
\end{aligned}
\label{eq:phi_ik_air}
\end{equation}
where $c_0$ is the speed of light $(3.0 \times 10^8~\mathrm{m/s})$, $f_0$ is the center frequency of the channel, $\Delta f$ is the subcarrier spacing $(312.5~\mathrm{kHz})$, $\lambda_k$ is the wavelength of subcarrier \#$k$, and $\tau_i$ is the propagation time of the direct wave, defined as $\tau_i \triangleq l_i/c_0$.

\begin{figure}[htbp]
  \centering
  \includegraphics[width=0.6\columnwidth]{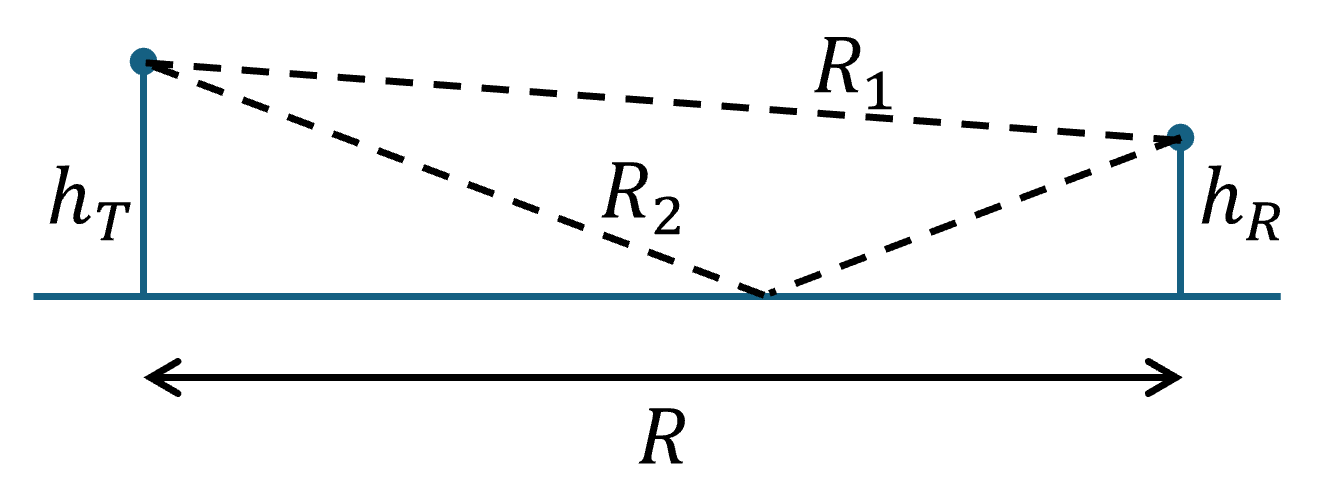}
  \caption{Two-ray ground reflection model ($R_1 = l_i$).}
  \label{fig:05}
\end{figure}

Even when a two-ray ground reflection model shown in Fig.~\ref{fig:05} is considered, the proportional relationship between $\varphi^{\mathrm{Air}}_{i,k}$ and subcarrier index $k$ is preserved when the horizontal distance $R$ is sufficiently larger than the antenna heights $h_T$ and $h_R$. 
In this case, the complex CSI can be approximated by
\begin{equation}
j \frac{4\pi h_T h_R}{\lambda_k R^2} e^{-j \frac{2\pi R_1}{\lambda_k}} ,
\label{eq:two_ray}
\end{equation} 
where $R_1$ denotes the direct-path length. 

\begin{figure}[htbp]
  \centering
  \includegraphics[width=0.7\columnwidth]{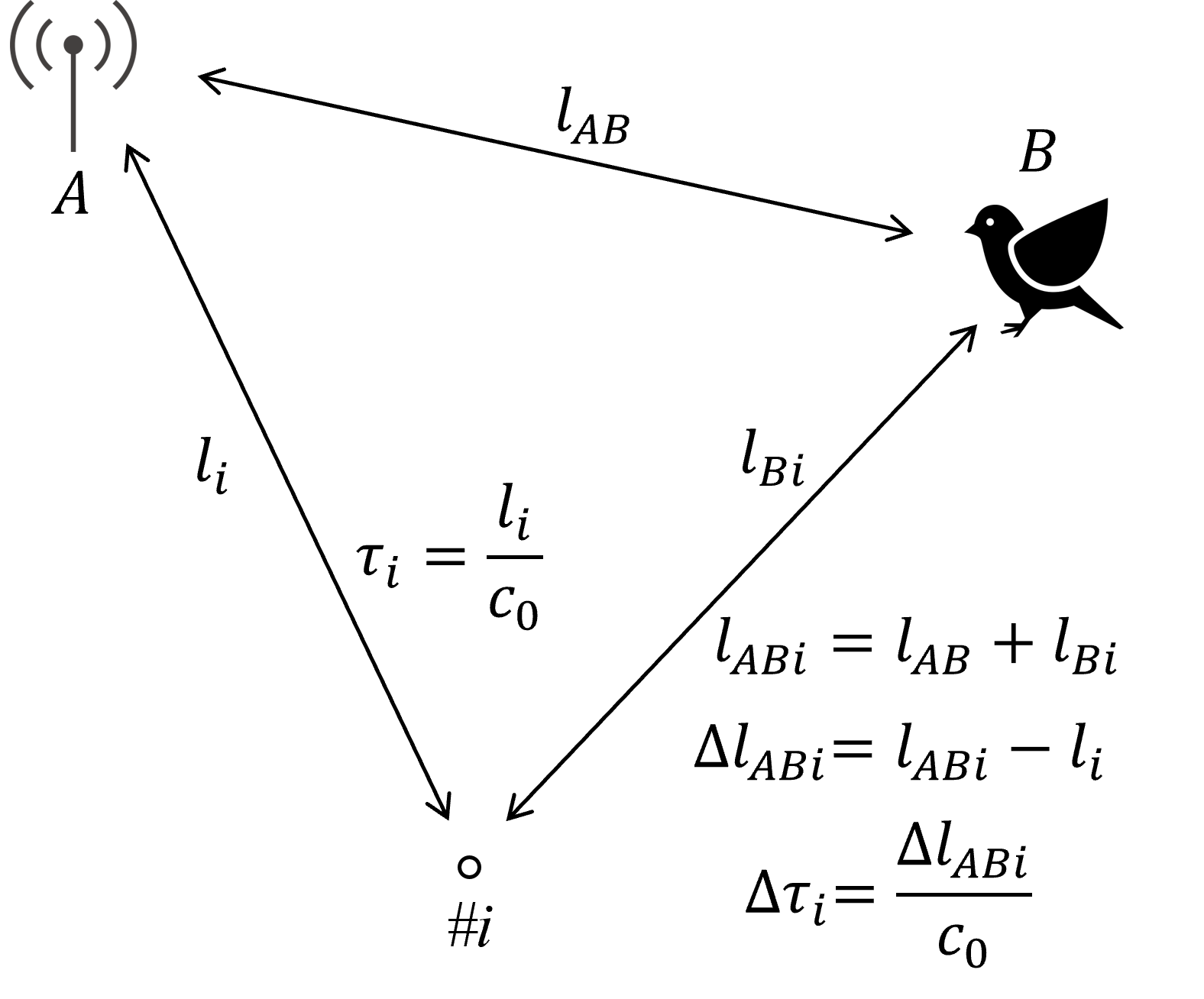}
  \caption{Reflection model.}
  \label{fig:06}
\end{figure}

When a wave reflected from an object exists, as shown in Fig.~\ref{fig:06}, the complex CSI and its phase component are expressed by \eqref{eq:hik} and \eqref{eq:phi_ik_airR_exact}, respectively. 
\begin{equation}
h_{i,k} = e^{-j \frac{2\pi}{\lambda_k} l_i} \left[ 1+b e^{-j \frac{2\pi}{\lambda_k}(l_{ABi}-l_i)} \right]
\label{eq:hik}
\end{equation}
\begin{equation}
\begin{aligned}
\angle h_{i,k} &\triangleq \varphi^{\mathrm{Air}}_{i,k} = -\frac{2\pi}{\lambda_k} l_i \\
&- \tan^{-1} \left( \frac{|b|\sin\left(\frac{2\pi}{\lambda_k}\Delta l_{ABi}-\angle b\right)}{1+|b|\cos\left(\frac{2\pi}{\lambda_k}\Delta l_{ABi}-\angle b\right)}\right)
\end{aligned}
\label{eq:phi_ik_airR_exact}
\end{equation}
where $|b|$ is the amplitude of the reflected wave from object $B$ normalized by the amplitude of the direct wave,  $\angle b$ is the phase angle of the reflection coefficient, and \[ \Delta l_{ABi} \triangleq l_{ABi} - l_i . \]

Assuming $|b| \ll 1$, the CSI phase can be approximated by 
\begin{equation}
\varphi^{\mathrm{Air}}_{i,k} \approx -\frac{2\pi}{\lambda_k} l_i-|b|\sin\left(\frac{2\pi}{\lambda_k} \Delta l_{ABi} - \angle b\right).
\label{eq:phi_ik_airR_approx}
\end{equation}

The $\Delta l_{ABi}$ is related to the position of object $B$ with respect to the Fresnel zone between antennas $A$ and \#$i$. 
Under the experimental environment described in Section~\ref{sec:me}, with a line-of-sight distance of 17~m and a channel frequency of 5600~MHz, the Fresnel radii shown in Table~\ref{tab:fresnel} are obtained. 
For example, when object $B$ is located within the 50th Fresnel zone, $\Delta l_{ABi}$ is at most approximately 25 wavelengths which is approximately 1.3~m ($300/5600\times25$). 
\begin{table}[htbp]
\caption{Radius of the $n$-th Fresnel Zone for a LoS Distance of 17~m at 5600~MHz}
\label{tab:fresnel}
\centering
\begin{tabular}{|c|c||c|c|}
\hline
\textbf{Order $n$} & \textbf{Radius $r_n$ (m)} & \textbf{Order $n$} & \textbf{Radius $r_n$ (m)} \\ \hline
1  & 0.477 & 20 & 2.151 \\ \hline
5  & 1.069 & 30 & 2.644 \\ \hline
10 & 1.515 & 40 & 3.065 \\ \hline
15 & 1.859 & 50 & 3.440 \\ \hline
\end{tabular}
\end{table}

From this scale relationship, defining the propagation delay difference as \[ \Delta \tau_i \triangleq \frac{\Delta l_{ABi}}{c_0}, \] and assuming \[ |2\pi k \Delta f \Delta \tau_i| \ll 1,\] the further approximation in \eqref{eq:phi_ik_airR_approx2} is obtained.
\begin{equation}
\begin{aligned}
\varphi^{\mathrm{Air}}_{i,k} &\approx -2\pi(f_0+k\Delta f)\tau_i \\
&\quad - |b|\sin(2\pi f_0 \Delta\tau_i - \angle b) \\
&\quad - 2\pi k\Delta f \Delta\tau_i |b| \cos(2\pi f_0 \Delta\tau_i - \angle b) \\
&= -\left[ 2\pi f_0 \tau_i +|b|\sin(2\pi f_0 \Delta\tau_i - \angle b) \right]\\
&\quad - 2\pi k\Delta f \left[\tau_i + \Delta\tau_i |b| \cos(2\pi f_0 \Delta\tau_i - \angle b) \right]
\end{aligned}
\label{eq:phi_ik_airR_approx2}
\end{equation}
Thus, the true CSI phase exhibits an approximate structure proportional to the subcarrier index $k$, similar to the device-induced phase offsets.

Focusing on this similarity, we consider the phase averaging of the symmetric subcarriers $\pm k$, defined as
\begin{equation}
\varphi_{i,\pm k} \triangleq \left( \frac{\varphi_{i,-k}+\varphi_{i,k}}{2}\right) \bmod 2\pi .
\label{eq:sym_avg}
\end{equation}

Using (\ref{eq:phi_ik}) and (\ref{eq:phi_ik_airR_approx2}), the following formulas can be derived from \eqref{eq:sym_avg}.
\begin{equation}
\varphi_{i,\pm k} =
\begin{cases}
\varphi_{i,0}, & \text{if \eqref{eq:no_flip_cond} is satisfied}, \\
\left( \varphi_{i,0} + \pi \right) \bmod 2\pi, & \text{otherwise},
\end{cases}
\label{eq:sym_avg_result}
\end{equation}
where 
\begin{equation}
\varphi_{i,0} \triangleq \left( \varphi_i - 2\pi f_0 \tau_i - |b| \sin(2\pi f_0 \Delta\tau_i - \angle b) \right) \bmod 2\pi
\label{eq:center_phase}
\end{equation}

\begin{equation}
\begin{aligned}
& \left| k \left( \Delta\varphi_i - 2\pi\Delta f \left( \tau_i + \Delta\tau_i |b| \cos(2\pi f_0 \Delta\tau_i - \angle b) \right) \right) \right| \\
&\qquad \approx \left| k (\Delta\varphi_i - 2\pi\Delta f \tau_i ) \right| < \frac{\pi}{2}
\end{aligned}
\label{eq:no_flip_cond}
\end{equation}
Thus, depending on condition \eqref{eq:no_flip_cond}, the phase average $\varphi_{i,\pm k}$ can be divided into two values with opposite sides.

On the other hand, $\pi$-shift does not occur when the phase variation is within $\pm\frac{\pi}{2}$ for all subcarriers in the channel. 
From this viewpoint, it is effective to calculate the inter-antenna phase differences before subcarrier averaging because the antenna-common phase rotations (clock-related phase errors) are canceled. 
Therefore, using antenna \#$j$ as the phase reference, the relative phase $\varphi_{ij,k}$ is defined as
\begin{equation}
\varphi_{ij,k} \triangleq (\varphi_{i,k} - \varphi_{j,k} ) \bmod 2\pi .
\label{eq:relative_phase}
\end{equation}

The compressed CSI phase is originally represented in this form, as shown in \eqref{eq:comp_csi}. 
The CSI phase difference is then averaged over the symmetric subcarriers $\pm k$ as
\begin{equation}
\varphi_{ij,\pm k} \triangleq \left( \frac{\varphi_{ij,-k} + \varphi_{ij,k}}{2} \right) \bmod 2\pi .
\label{eq:relative_avg}
\end{equation}

When condition \eqref{eq:relative_cond} is satisfied, then \eqref{eq:relative_avg} is given by \eqref{eq:relative_center}:
\begin{equation}
\left| k ( \Delta\Phi_{ij} - 2\pi\Delta f\tau_{ij} )\right| < \frac{\pi}{2}
\label{eq:relative_cond}
\end{equation}

\begin{equation}
\varphi_{ij,\pm k} = \varphi_{ij,0} \triangleq ( \Phi_{ij} + \varphi^{\mathrm{Air}}_{ij,0} ) \bmod 2\pi
\label{eq:relative_center}
\end{equation}

where
\begin{equation}
\begin{aligned}
\Phi_{ij} &\triangleq \Phi_i - \Phi_j, \\
\Delta\Phi_{ij} &\triangleq \Delta\Phi_i-\Delta\Phi_j, \text{and} \\
\tau_{ij} &\triangleq \frac{l_{ij}}{c_0} \triangleq \frac{l_i-l_j}{c_0}.
\end{aligned}
\label{eq:relative_def}
\end{equation}

The true relative CSI phase is approximately expressed as
\begin{equation}
\begin{aligned}
\varphi^{\mathrm{Air}}_{ij,0} &\triangleq \varphi^{\mathrm{Air}}_{i,0} - \varphi^{\mathrm{Air}}_{j,0} \\
&\approx -2\pi f_0\tau_{ij} \\
&\quad - |b| \left[ \sin(2\pi f_0\Delta\tau_i-\angle b) - \sin(2\pi f_0\Delta\tau_j-\angle b) \right].
\end{aligned}
\label{eq:relative_air}
\end{equation}
Here, the reflection wave coefficient $b$ is assumed to be identical for antennas \#$i$ and \#$j$.

The proportional coefficient $\Delta\Phi_{ij}$ of the phase offset originates from differences in RF cable length and filter characteristics between antenna paths. 
The direct-path difference $l_{ij}$ depends on the antenna spacing between antennas \#$i$ and \#$j$. 
The proportional coefficient $\Delta\Phi_{ij}$ is rewritten using the equivalent RF-chain distance difference $\Delta\Lambda_{ij}$ as
\begin{equation}
\Delta\Phi_{ij} = \frac{2\pi\Delta f}{c_0} \Delta\Lambda_{ij}.
\label{eq:equivalent_distance}
\end{equation}

In this case, phase inversion errors do not occur if
\begin{equation}
|k(\Delta\Lambda_{ij}-l_{ij})| < \frac{c_0}{4\Delta f} = 240~(\mathrm{m}).
\label{eq:no_flip_distance}
\end{equation}
Assuming the maximum subcarrier index is $k=28$, the phase mismatch caused by differences between antenna paths is allowable up to approximately 8.5~m in terms of equivalent distance $|\Delta\Lambda_{ij}-l_{ij}|$. 
This condition is expected to be satisfied in commercial devices, considering the typical antenna spacing and RF cable length differences.

\subsection{Combining and Differentiation of the Subcarrier Phase}

From \eqref{eq:relative_center}, the averaged phase difference of the symmetric subcarriers $\pm k$, $\varphi_{ij,\pm k}$, is independent of the subcarrier index $k$ and is equal to the center phase difference $\varphi_{ij,0}$. 
Using this property, the phase differences $\varphi_{ij,k}$ over all the subcarriers are combined by averaging. 
This operation removes the proportional component $\Delta\Phi_{ij}$ of the phase offset contained in $\varphi_{ij,k}$, while also providing a processing gain that suppresses the thermal and quantization noise associated with the CSI phase differences. 
These noise components can be regarded as random, additive, and zero-mean for each subcarrier, and are therefore mutually canceled by averaging. 
By contrast, the true CSI phase-difference component $\varphi^{\mathrm{Air}}_{ij,0}$, whose sign is aligned by symmetric subcarrier combining, coherently accumulates, thereby improving the SNR. 
However, the constant phase offset component $\Phi_{ij}$ remains and is removed by temporal differentiation, as described below.

Let $\varphi_{ij,k,n}$ denote the CSI phase difference measured at discrete time index $n$. 
In a low-multipath environment, the averaged phase difference $\varphi_{ij,\pm k,n}$ of symmetric subcarriers $\pm k$ becomes approximately identical to the center phase $\varphi_{ij,0,n}$. 
However, when the center phase is located near the phase-wrapping boundary, phase wrapping may occur for some subcarriers because of noise. 
Therefore, when averaging all the symmetric subcarriers, double phase differences are calculated by subtracting the reference phase difference $\varphi_{ij,\pm 1,n}$ from $\varphi_{ij,\pm k,n}$, followed by a modulo operation with $2\pi$. 
After averaging these double phase differences, $\varphi_{ij,\pm 1,n}$ is added back, and the result is again taken modulo $2\pi$ to avoid phase-wrapping errors. 
As an example for the compressed CSI phase with 20MHz channel width, this procedure is expressed as
\begin{equation}
\begin{aligned}
\bar{\varphi}_{ij,n} = & \Biggl[ \frac{1}{26} \sum_{\substack{k=2 \\ k\ne 7,\,21}}^{28} \left\{ \left( \varphi_{ij,\pm k,n} - \varphi_{ij,\pm 1,n} \right) \bmod 2\pi \right\} \\
& \qquad + \varphi_{ij,\pm 1,n} \Biggr] \bmod 2\pi .
\end{aligned}
\label{eq:phase_combining}
\end{equation}

The expected value of $\bar{\varphi}_{ij,n}$ in \eqref{eq:phase_combining} is $\varphi_{ij,0}$ in \eqref{eq:relative_center}. 
The noise associated with $\bar{\varphi}_{ij,n}$ originates from the thermal and quantization noises contained in the averaged symmetric subcarrier phase differences $\varphi_{ij,\pm k}$. The mean squared error is reduced to $1/26$ of the original averaged quantities. 
Compared with the noise associated with a single-subcarrier phase difference $\varphi_{ij,k}$, it is reduced to $1/52$.

As shown in \eqref{eq:relative_center}, $\bar{\varphi}_{ij,n}$ includes the constant phase-offset term $\Phi_{ij}$. 
This term is caused by differences in the RF chains of the antennas, such as RF cable lengths and analog filter characteristics, and varies slowly with time. 
Therefore, the constant phase-offset term is removed by taking the temporal difference of $\bar{\varphi}_{ij,n}$. 
Specifically, the phase-difference variation is quantified by the frequency-deviation metric $f_{ij,n}$ defined as
\begin{equation}
\Delta\varphi_{ij,n} = \left( \bar{\varphi}_{ij,n+1} - \bar{\varphi}_{ij,n} \right) \bmod 2\pi ,
\label{eq:phase_difference}
\end{equation}
and
\begin{equation}
f_{ij,n} = \frac{1}{2\pi} \frac{\Delta\varphi_{ij,n}}{\Delta t_n}, \label{eq:phase_rate}
\end{equation}
where $\Delta t_n$ is the time interval between discrete time indices $n$ and $n+1$. 
The quantity $f_{ij,n}$ represents the temporal rate of phase variation expressed in units of frequency (Hz).
Motion detection is performed on the basis of the magnitude of this metric.

\section{Evaluation of the Proposed Method Using Measured Orchard Data}
\label{sec:eval}
\subsection{Application of the Proposed Method}

\begin{figure}[htbp]
\centering
\includegraphics[width=0.5\columnwidth]{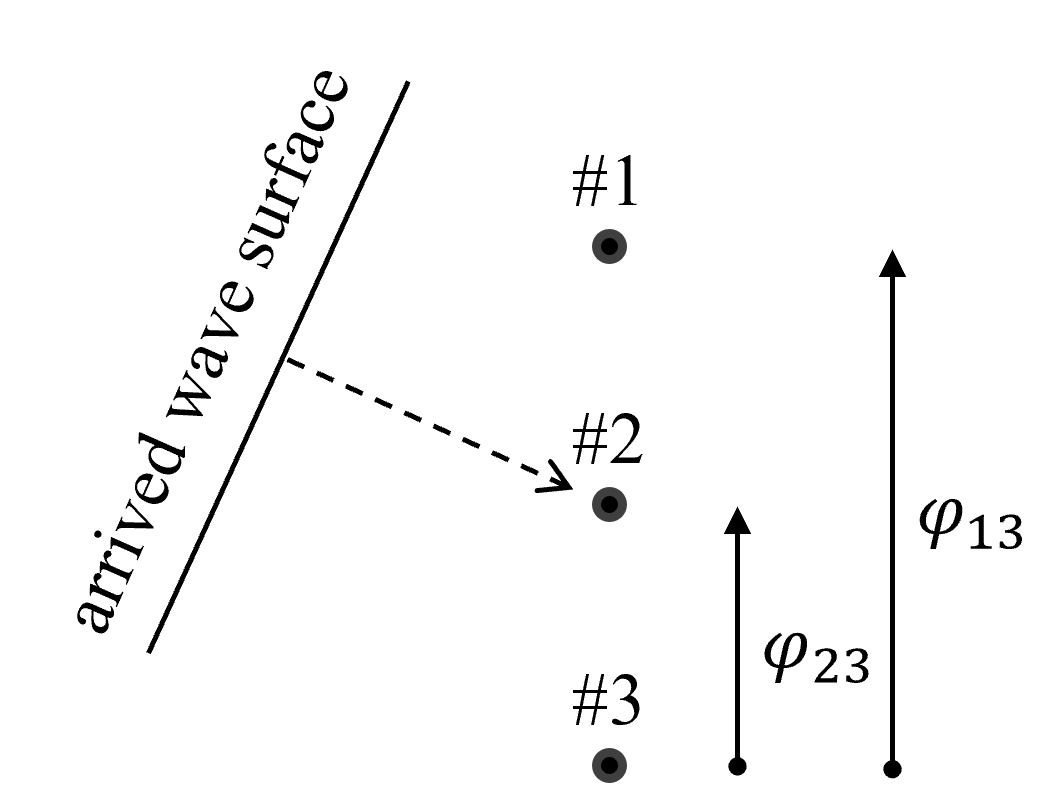}
\caption{Application to 3-Antenna System.}
\label{fig:07}
\end{figure}

Using the measurement configuration shown in Table \ref{tab:spec}, two center phase displacements, $\Delta\varphi_{13,n}$ and $\Delta\varphi_{23,n}$, are obtained from a single AP-STA pair. 
These correspond to $-\varphi_{11}$ and $-\varphi_{21}$ in \eqref{eq:comp_csi}, respectively. 
As shown in Fig.~\ref{fig:07}, the signs of the inter-antenna phase differences are aligned according to the arrangement of AP antenna positions \#1-\#3 so that the phase variations caused by incoming waves have the same direction. 
The aligned phase differences are then combined by signed averaging to further suppress noise. 
Specifically, the following quantities are calculated and their variations during object intrusion are evaluated:
\begin{equation}
\Delta\varphi_{123,n} \triangleq \frac{\Delta\varphi_{13,n} + \Delta\varphi_{23,n}}{2},
\label{eq:phase123}
\end{equation}
\begin{equation}
f_{123,n} \triangleq \frac{1}{2\pi} \frac{\Delta\varphi_{123,n}}{\Delta t_n}.
\label{eq:f123}
\end{equation}

The frequency deviation $f_{123,n}$ is calculated independently for STA\#1 and STA\#2 and plotted for each 10-minute interval. 
The 10-minute segmentation allows the temporal behavior before and after object movement to be visually confirmed. 
To improve graph readability while appropriately reducing the number of plotted points, the maximum and minimum frequency displacements within each one-second interval are calculated, and their difference (range) is plotted every second. 
Since approximately 32 Compressed Beamforming frames per second were obtained in the measurements, the detectable maximum and minimum frequency deviations are approximately $\pm 16$~Hz. 
Accordingly, the detectable range of frequency displacement is approximately 32~Hz. For reference, the measured wind-speed data are also presented.

\subsection{Frequency Deviations During SUV Intrusion}

\begin{figure}[tp]
 \centering
 \subfloat[Superimposed video frames illusrating SUV intrusion.]{ 
  \includegraphics[width=\columnwidth]{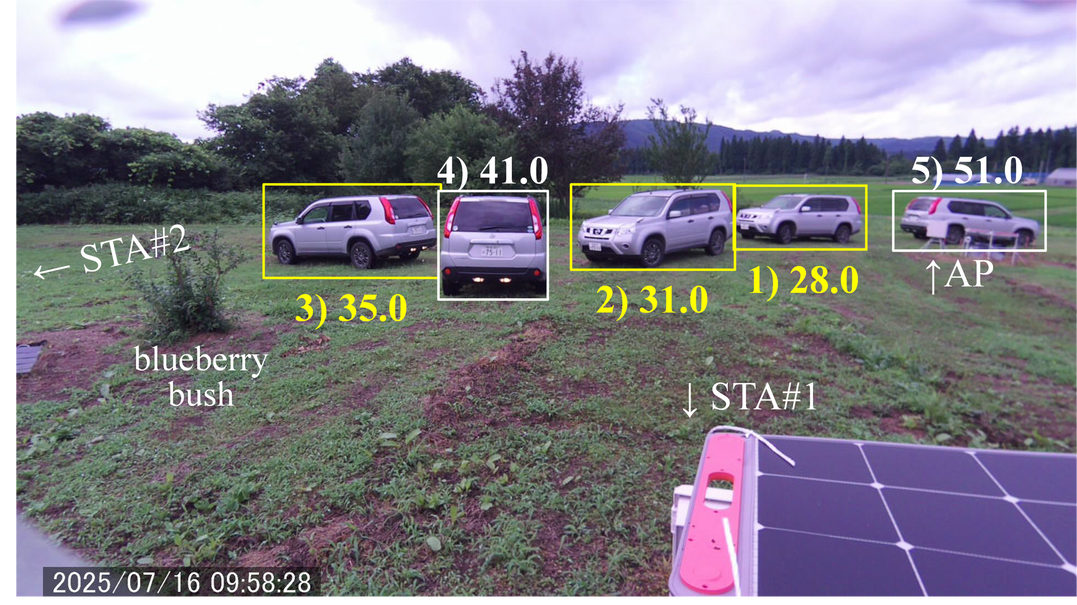}
  \label{fig:08a}
 } \\
 \subfloat[STA\#1 frequency deviation from 09:50 to 10:00.]{
  \includegraphics[width=\columnwidth]{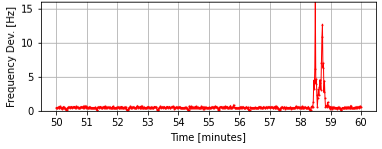}
  \label{fig:08b}
 } \\
  \subfloat[STA\#2 frequency deviation from 09:50 to 10:00.]{
  \includegraphics[width=\columnwidth]{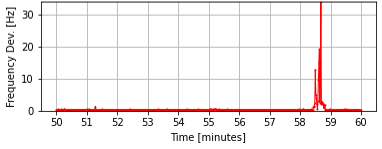}
  \label{fig:08c}
 } \\
  \subfloat[Wind speed  from 09:50 to 10:00.]{
  \includegraphics[width=\columnwidth]{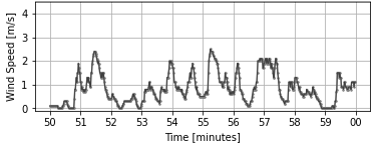}
  \label{fig:08d}
 }
 \caption{Motion of SUV Intrusion at 09:58.}
 \label{fig:08}
\end{figure}

As a reference example representing the largest motion-induced deviation observed during the measurement period, this subsection presents the case of vehicle intrusion rather than bird intrusion. 
The largest deviation during the evaluated measurement period was observed at 09:58, when a compact SUV entered the field area, turned around (switchback maneuver), and departed. 
This situation is shown in Fig.~\ref{fig:08}. 
The video was recorded from the camera near the STA\#1 side, where the AP is located near the center-right in the video frame of Fig.~\ref{fig:08a}. 
To visualize the motion of the object, multiple video frames are superimposed within the image, and their timestamps (in seconds) are indicated together with sequential frame numbers. 
The background image corresponds to frame number 1). 
The same representation is used in Figs.~\ref{fig:09}-\ref{fig:11}.

In this case, the SUV almost touched LoS\#2 (see Fig.~\ref{fig:03}) twice, between superimposed frames 2) and 3) and frame 4), resulting in a frequency deviation exceeding 30~Hz at STA\#2. 
From the viewpoint of LoS\#1 (see Fig.~\ref{fig:03}), the closest approach occurred when the SUV passed near the AP. 
As estimated from Fig.~\ref{fig:08a}, the minimum distance was approximately 2~m from LoS\#1. 
Consequently, the maximum frequency deviation observed for LoS\#1 was smaller than that for LoS\#2 and remained approximately 16~Hz.

\subsection{Frequency Deviations During Bird Intrusion}
\paragraph*{Case 1: Direct Flight of a Crow to the Bush}

\begin{figure}[tp]
 \centering
 \subfloat[Superimposed video frames illusrating case 1.]{
  \includegraphics[width=\columnwidth]{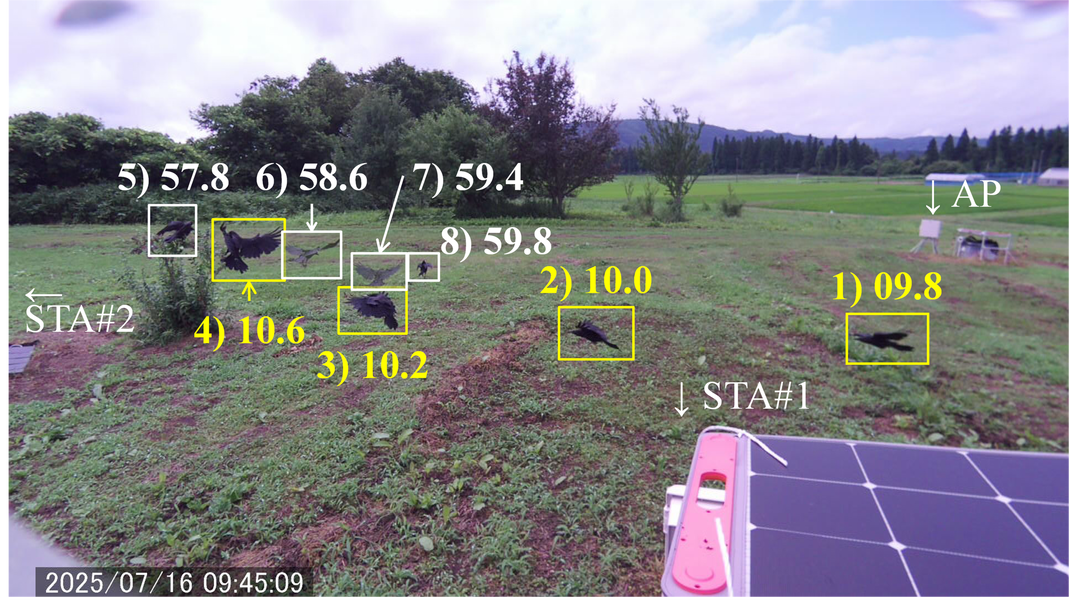}
  \label{fig:09a}
 } \\
 \subfloat[STA\#1 frequency deviation from 09:40 to 09:50.]{
  \includegraphics[width=\columnwidth]{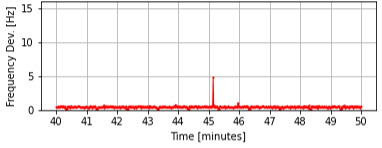}
  \label{fig:09b}
 } \\
  \subfloat[STA\#2 frequency deviation from 09:40 to 09:50.]{
  \includegraphics[width=\columnwidth]{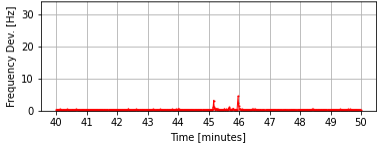}
  \label{fig:09c}
 } \\
  \subfloat[Wind speed  from 09:40 to 09:50.]{
  \includegraphics[width=\columnwidth]{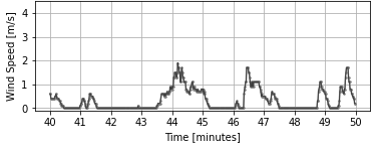}
  \label{fig:09d}
 }
 \caption{Case 1: Flying Directly to the Bush at 09:45.}
 \label{fig:09}
\end{figure}

This example corresponds to one of the cases introduced in Section~\ref{ssec:bird-speed}. 
At 09:45, a crow flew directly toward the blueberry bush and remained there for approximately one minute while pecking the fruit. 
Figure~\ref{fig:09} shows the measurement results obtained using the proposed method together with the corresponding scene images. 
In the photographs, frames 1)-4) correspond to the crow approaching the bush, whereas frames 5)-8) show the crow leaving the bush.

During the intrusion, the crow crossed LoS\#1, and a frequency deviation of approximately 5~Hz was detected at STA\#1. 
Additional variations were detected at STA\#2 during the latter half of 09:45:10, when the crow arrived at the blueberry bush, and again near 09:46, when it departed. 
Although the departure path of the crow was visually estimated to be approximately 5~m away from LoS\#1, a slight deviation was still observable at STA\#1.

\paragraph*{Case 2: Low-Altitude Crow Flyby}

\begin{figure}[tp]
 \centering
 \subfloat[Superimposed video frames illusrating case 2.]{
  \includegraphics[width=\columnwidth]{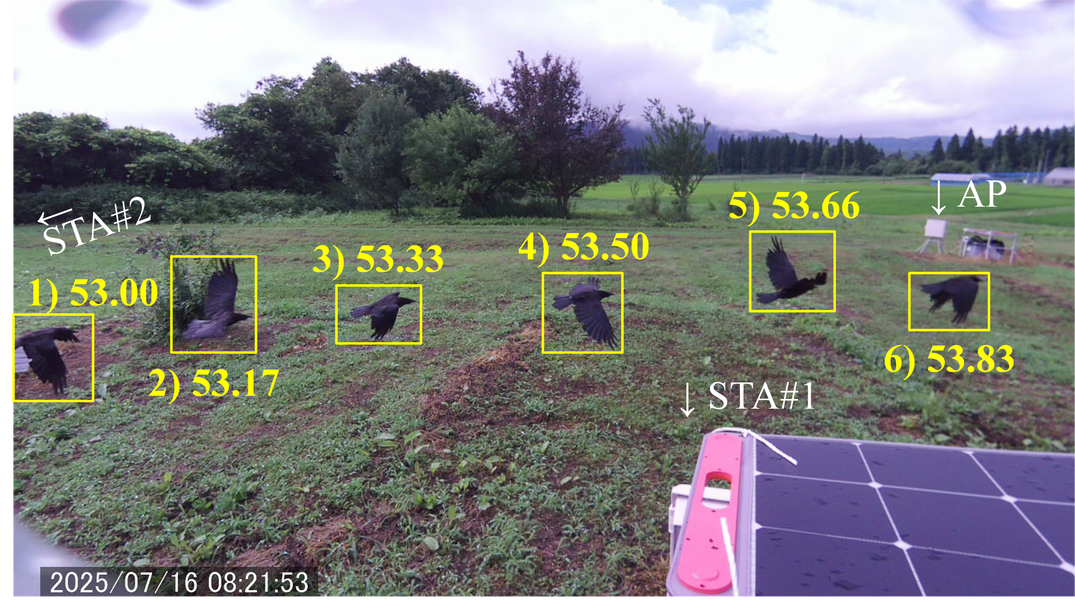}
  \label{fig:10a}
 } \\
 \subfloat[STA\#1 frequency deviation from 08:20 to 08:30.]{
  \includegraphics[width=\columnwidth]{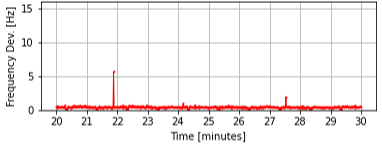}
  \label{fig:10b}
 } \\
  \subfloat[STA\#2 frequency deviation from 08:20 to 08:30.]{
  \includegraphics[width=\columnwidth]{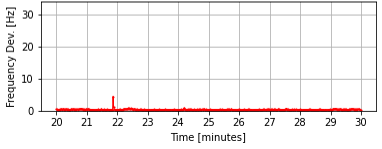}
  \label{fig:10c}
 } \\
  \subfloat[Wind speed  from 08:20 to 08:30.]{
  \includegraphics[width=\columnwidth]{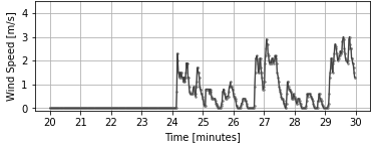}
  \label{fig:10d}
 }
 \caption{Case 2: Low-Altitude Crow Flyby at 08:21.}
 \label{fig:10}
\end{figure}

Figure~\ref{fig:10} shows a case in which a crow approached the monitored area at low altitude and crossed LoS\#1 at 08:21:53. 
Although the crow was approximately 4~m away from LoS\#2, a deviation of approximately 5~Hz was detected at STA\#2. 
Considering the Fresnel zone radii listed in Table~\ref{tab:fresnel}, this result confirms that the flyby of a single crow could be detected even beyond the 50th Fresnel zone.

In addition, a deviation of approximately 2.5~Hz was detected for STA\#1 at approximately 08:27:30.
However, no flying objects appeared in the recorded video at that time, although bird calls were captured in the audio. 
Considering the wind speed conditions, it is unlikely that the deviation was caused by the wind-induced vegetation motion. 
Together with the fact that the observed deviation pattern was similar to that of Case~3 described below, it is likely that a bird flew through a blind spot outside the camera field of view.

On the other hand, a very slight deviation observed at STA\#1 at approximately 08:24:10 coincided with the timing at which a starling flew from the camera side toward the blueberry bush. 
According to the wind speed data collected during the same period, the wind speed rapidly increased from calm conditions to approximately 2.5~m/s. 
At the initial stage of the investigation, the deviation was suspected to have been caused by sudden gust. 
However, the recorded video did not show any abrupt vegetation motion, such as rapid shaking of leaves or branches.

\paragraph*{Case 3: Bird Flying Overhead}

\begin{figure}[tp]
 \centering
 \subfloat[Superimposed video frames illusrating case 3.]{
  \includegraphics[width=\columnwidth]{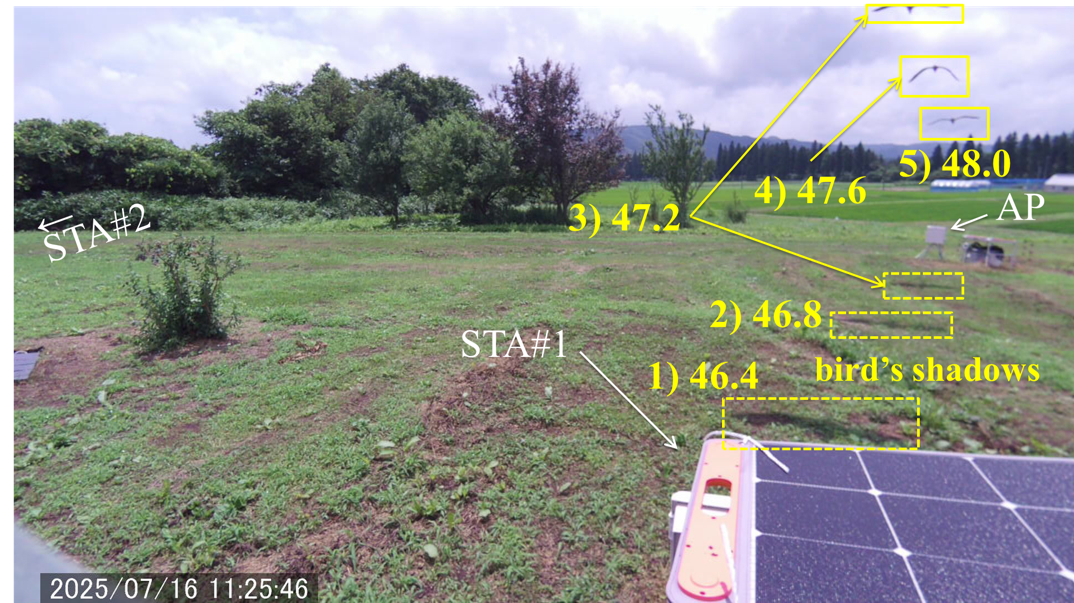}
  \label{fig:11a}
 } \\
 \subfloat[STA\#1 frequency deviation from 11:20 to 11:30.]{
  \includegraphics[width=\columnwidth]{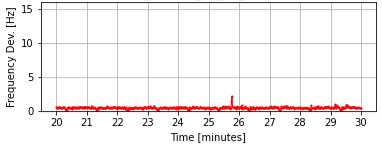}
  \label{fig:11b}
 } \\
 \subfloat[STA\#2 frequency deviation from 11:20 to 11:30.]{
  \includegraphics[width=\columnwidth]{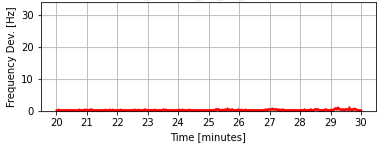}
  \label{fig:11c}
 } \\
 \subfloat[Wind speed  from 11:20 to 11:30.]{
  \includegraphics[width=\columnwidth]{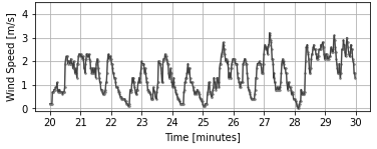}
  \label{fig:11d}
 }
 \caption{Case 3: Bird Flying Overhead at 11:25.}
 \label{fig:11}
\end{figure}

This event was discovered using the proposed method and corresponds to an overhead bird flight that had initially been overlooked during manual inspection of the video recordings alone. 
This situation is shown in Fig.~\ref{fig:11}. 
The bird was estimated to be flying at an altitude of approximately 4~m while moving roughly along LoS\#1, and a deviation of approximately 2.5~Hz was detected at STA\#1. 
In contrast, no significant variation was observed at STA\#2, confirming that the flight path was less favorable for detection from the viewpoint of LoS\#2 than that of LoS\#1.

\subsection{Influence of Environmental and Weather Variations}

In the cases shown in Figs.~\ref{fig:08}-\ref{fig:11}, intermittent winds with speeds ranging from 0 to 3~m/s was observed. 
However, no noticeable influence of wind-induced vegetation motion was observed in the frequency-deviation plots.

As described in Section~\ref{ssec:ms}, light rainfall was recorded at a nearby meteorological observatory approximately two hours before the measurements and again between 09:50 and 10:00 during the measurements. 
During these periods, the frequency deviations shown in Figs.~\ref{fig:08b} and \ref{fig:08c} did not exhibit any unusual behavior except for the SUV intrusion event. 

According to \cite{naohiro13}, rainfall corresponding to 0.5~mm per 10~min (3~mm/h) typically consists of raindrops with diameters up to approximately 2~mm. 
Furthermore, assuming raindrop diameters below 1~mm, the terminal velocity near ground level is generally less than 4~m/s \cite{gunn1949}.

The events shown in Figs.~\ref{fig:08} and \ref{fig:09} occurred between 09:45 and 09:59, whereas the event shown in Fig.~\ref{fig:10} occurred during 08:21. 
In all these figures, water droplets remained visible on the camera hood located at the upper left of the Figs.~\ref{fig:08a}, \ref{fig:09a} and \ref{fig:10a}. 
Furthermore, water droplets are visible on the solar panel surface, as shown in Fig.~\ref{fig:10a}. 
In contrast, the event shown in Fig.~\ref{fig:11} occurred before 11:30, when the water droplets had disappeared, and sunlight was confirmed from the bird shadow and reflections on the solar panel on Fig.~\ref{fig:11a}.

Despite these environmental variations, their influence on the proposed sensing method was found to be extremely limited. 
This is considered to be because vegetation motion and light rainfall droplets produce much weaker reflections than medium-sized birds such as crows, while their movement speeds were generally below approximately 3~m/s.

\section{Conclusion}
\label{sec:conclude}
This paper proposes a motion-detection method that suppresses device-induced phase errors, quantization noise, and thermal noise while avoiding failures in phase-unwrapping processing by exploiting the property that the averaged CSI phase of symmetric subcarriers converges to the center channel phase in low-multipath environments. 
The proposed metric is obtained by averaging the deviations from the center phase difference and subsequently adding them to avoid potential errors due to phase wrapping. 
Both the theoretical analysis and experimental results demonstrated that the proposed method is highly effective in low-multipath environments. 
Particularly noteworthy is that bird flights several meters away from the direct LoS path were successfully detected.
In the measured orchard environment, the noise level remained below approximately 1~Hz, even under wind conditions of approximately 3~m/s, confirming that the influence of environmental variations was limited. 
Although compressed CSI was used in the experiments, the proposed method is also applicable to ordinary CSI measurements.

While this study focused on orchard monitoring, the low-multipath environments considered in this work can also arise in suburban monitoring, infrastructure monitoring, and future beamforming-based ISAC systems. 
Therefore, the findings obtained in this study are expected to contribute to the fundamental understanding of CSI-based sensing in low-multipath environments.

\section*{Acknowledgment}

This work was supported by JSPS KAKENHI Grant Number JP23K03870.
Generative AI tools (specifically, OpenAI ChatGPT based on GPT-5.5) were used during the manuscript preparation process for English editing and stylistic refinement. 
The authors take full responsibility for all technical content and conclusions.

\begin{IEEEbiography}[{\includegraphics[width=1in,height=1.25in,clip,keepaspectratio]{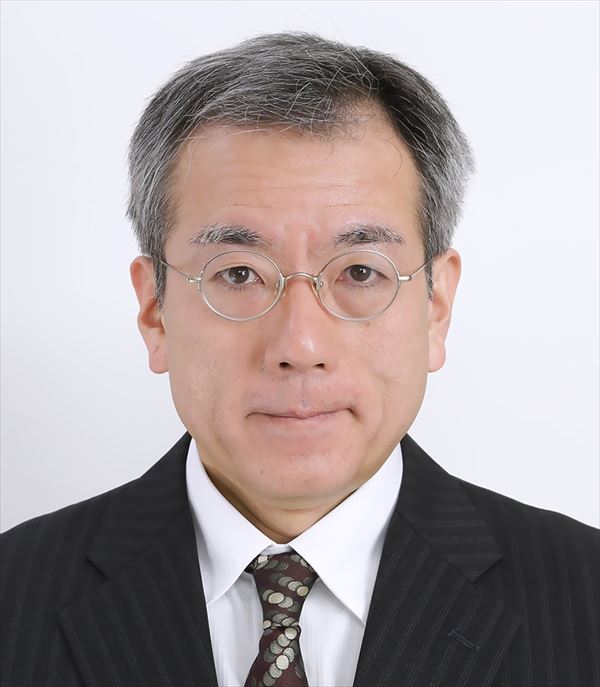}}]
{Toshinori Suzuki} (M'00) received the B.E., M.E., and Ph.D. degrees in electrical engineering from Tohoku University, Sendai, Japan, in 1987, 1989, and 2005, respectively.

In 1989, he joined Kokusai Denshin Denwa Co., Ltd (KDD, currently KDDI). After spending two years at the KDD Yamaguchi Satellite Communication Station, he was with KDDI R\&D Laboratories, Inc., from 1991 to 2011, where he was mainly engaged in the research and development of mobile communication technologies, WiFi applications, and security systems.
Since 2011, he has been a Professor with the Department of Information Technology Engineering, Tohoku Gakuin University. 
His research interests include radio communication systems, integrated sensing and communication (ISAC) technologies, and GNSS.

Prof. Suzuki is a member of the Institute of Electronics, Information and Communication Engineers (IEICE), where he was elevated to the grade of Fellow in 2019. 
He has received the IEICE Communications Society Distinguished Contributions Award in 2004 and 2018, the IEICE Best Paper Award in 2005, the Research Encouragement Award from the Minoru Ishida Foundation in 2012, and the Invention Encouragement Award from the Japan Institute of Invention and Innovation in 2016.
\end{IEEEbiography}

\begin{IEEEbiography}[{\includegraphics[width=1in,height=1.25in,clip,keepaspectratio]{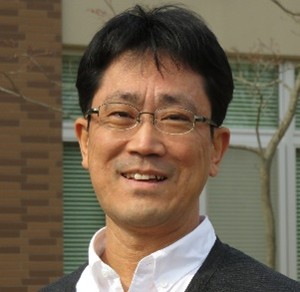}}]
{Shin-ichiro OGURA} received B.Agr., M.S. in Agr. and Ph.D. degrees in Agricultural Science from Tohoku University, Sendai, Japan, in 1992, 1994 and 1997, respectively.

He worked as a post doctoral fellow of the Science and Technology Promotion Agency in the Livestock Division of the Chugoku Agricultural Experiment Station, Ministry of Agriculture, Forestry and Fisheries in 1997. 
After spending 1 year, he belonged to the Faculty of Agriculture, Miyazaki University from 1998 to 2003, as an Assistant Professor. 
From 2003, he belonged to the Graduate School of Agricultural Science, Tohoku University, as an Associate Professor. 
Since 2016, he has been a Professor of the Graduate School of Agricultural Science, Tohoku University. 
His research focuses on the plant-animal interactions in grassland ecosystems, Forage production and utilization by ruminants, and wildlife ecology and measures to mitigate wildlife damage in agriculture.

Prof. Ogura is a member of the Japanese Society for Animal Behavior and Management, Japanese Society of Animal Science, Japanese Society of Grassland Science, etc.
\end{IEEEbiography}



\begin{IEEEbiography}[{\includegraphics[width=1in,height=1.25in,clip,keepaspectratio]{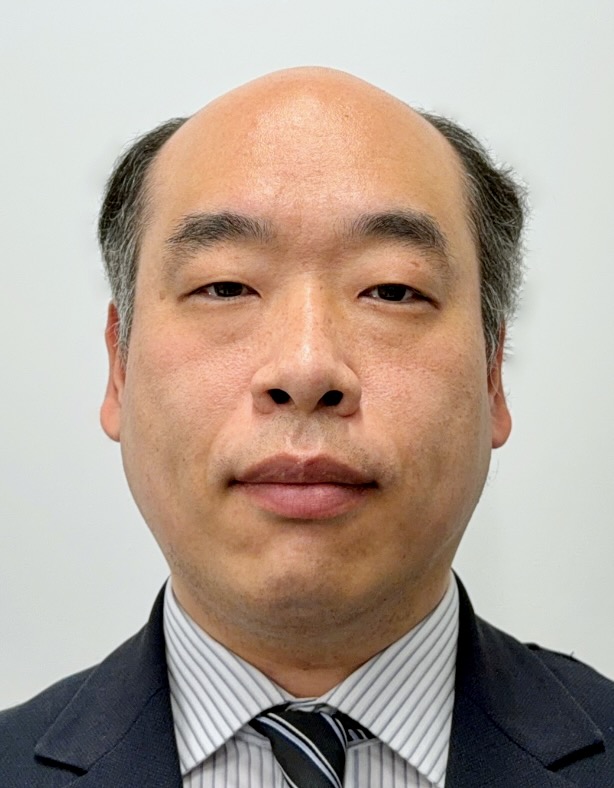}}]
{Yu Morishima} (M'14)
received his B.S., M.S., and Ph.D. degrees in Engineering from Osaka City University in 2009, 2011, and 2014, respectively. 

From 2014 to 2018, he served as an Assistant Professor and from 2018 to 2019 as a Lecturer at National Institute of Technology, Suzuka College. 
Since 2019, he has been a Lecturer at Tohoku Gakuin University. 
His research interests include error-correction codes, information theory, and wireless communication.

Dr. Morishima is a member of  the Institute of Electronics, Information and Communication Engineers (IEICE) and received the Distinguished Contribution Award of the IEICE Engineering Sciences Society (ESS) in 2023.
\end{IEEEbiography}

\begin{IEEEbiography}[{\includegraphics[width=1in,height=1.25in,clip,keepaspectratio]{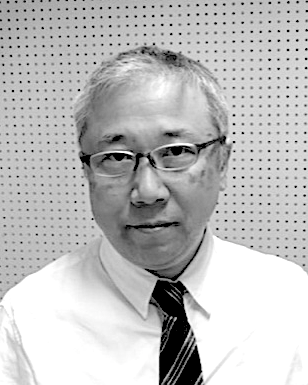}}]
{HIROSHI MATSUURA}, Ph.D., joined the Central Research Laboratory of Sumitomo Cement Co., Ltd. in 1987, where he was engaged in the development of optical communication devices.

In 1998, he joined Furukawa Electric Co., Ltd., where he conducted research on optical fiber devices and lasers. 
Since 2011, he has been a Professor in the Department of Mechanical Engineering and Intelligent Systems, Tohoku Gakuin University.
His current work includes a MIC-supported FORWARD project focusing on harmful-animal sensing systems using LoRa and Wi-Fi technologies in off-grid areas. 
He also develops LoRaMap, an Android map application used with 920 MHz-band LoRa devices to help prevent users from getting lost.

Prof. Matsuura is a member of the Institute of Electronics, Information and Communication Engineers (IEICE) and the Japan Society for Precision Engineering (JSPE).
\end{IEEEbiography}

\EOD
\end{document}